\documentclass[a4paper,conference]{IEEEtran}

\usepackage{cite}

\usepackage{tikz}
\usetikzlibrary{dsp,chains}

\usepackage{color}
\definecolor{green}{rgb}{0,0.4,0.05}
\definecolor{red}{rgb}{0.8,0,0}
\usepackage{psfrag}

\usepackage{pgfplots}
\usepackage{authblk}
\usepackage{amsbsy}
\usepackage{amsmath}
\usepackage{fixmath}
\usepackage{amssymb}

\usepackage{graphicx}

\usepackage{algorithm}
\usepackage{algpseudocode}
\newcommand{\vect}[1]{\mathbf{#1}}
\newcommand{\matt}[1]{\mathbf{#1}}
\newcommand{\jim}{\mathrm{j}\,}

\newcommand{\T}{\operatorname{\mathrm{T}}}
\newcommand{\Q}{\operatorname{\mathcal{Q}}}
\newcommand{\He}{\text{H}}
\newcommand{\diag}{\text{diag}}

\begin{document}
\title{Massive MIMO Downlink 1-Bit Precoding with Linear Programming for PSK Signaling}

\author[1]{Hela~Jedda}
\author[2]{Amine~Mezghani}
\author[1,3]{Josef~A.~Nossek}
\author[2]{A.~Lee~Swindlehurst}
\affil[1]{Technical University of Munich, 80290 Munich, Germany}
\affil[2]{University of California, Irvine, Irvine, CA 92697, USA}
\affil[3]{Federal University of Cear\'a, Fortaleza, Brazil}
\affil[ ] {Email: hela.jedda@tum.de, amezghan@uci.edu, josef.a.nossek@tum.de}

\maketitle
\begin{abstract}
Quantized massive multiple-input-multiple-output (MIMO) systems are gaining more interest due to their power efficiency. We present a new precoding technique to mitigate the multi-user interference and the quantization distortions in a downlink multi-user (MU) multiple-input-single-output (MISO) system with 1-bit quantization at the transmitter. This work is restricted to PSK modulation schemes. The transmit signal vector is optimized for every desired received vector taking into account the 1-bit quantization. The optimization is based on maximizing the safety margin to the decision thresholds of the PSK modulation. Simulation results show a significant gain in terms of the uncoded bit-error-ratio (BER) compared to the existing linear precoding techniques.
\end{abstract}

\IEEEpeerreviewmaketitle
\section{Introduction}
\label{sec:intro}

For the next generation of mobile communication, where massive multiple-input-multiple-output (MIMO) systems are foreseen as one of the key technologies, power consumption is a crucial concern due to the deployment of a large number of antennas and hence the corresponding RF chains. Green communication aims at minimizing the carbon dioxide emissions while guaranteeing the quality of service. One aspect consists in reducing the hardware power consumption mainly of the power amplifiers (PAs) that are considered as the most power hungry devices at the transmitter side \cite{Blume2010, Chen2010}. When the PAs are run in the saturation region high power efficiency is achieved. However, in the saturation region strong nonlinear distortions are introduced to the signals. To avoid the PA distortions when run in the saturation region, we resort to PA input signals of constant envelope. Constant envelope signals have the property of constant magnitude. Thus, the magnitude distortions are omitted.

In this spirit, the deployment of 1-bit digital-to-analog converters (DACs) at the transmitter ensures on the one hand the property of constant envelope signals at the input of the PA. On the other hand the power consumption of the DAC itself is minimized. Therefore, the power efficiency goal is achieved twice: power efficient PA due to the constant envelope signals and less power consuming DACs due to the low resolution. The use of 1-bit DACs is also beneficial in terms of reduced cost and circuit area and can further simplify the surrounding RF circuitry due to the relaxed linearity constraint, leading to very efficient hardware implementations \cite{Chen_2015}. However, the coarse quantization causes nonlinear distortions that degrade the performance. Therefore, mitigating the quantization distortions has to be considered in the precoding task in multi-user (MU) MIMO systems.

The contribution in \cite{Mezghani2009} is an early work that addressed the precoding task with low resolution quantization at the transmitter. The authors in \cite{Usman2016} introduced another linear precoder that could slightly improve the system performance. The proposed precoder is designed based on an iterative algorithm since no closed form expression can be obtained. Theoretical analysis on the achievable rate in systems with 1-bit transmitters were investigated in \cite{Kakkavas2016, Saxena2016, Yongzhi2016}. Nonlinear precoding techniques in this context were introduced in \cite{JeddaSAM2016}. The authors presented a symbol-wise precoding technique based on a minimum bit error ratio (MBER) criterion and made use of the box norm ($\ell_{\infty}$) to relax the 1-bit constraint. In \cite{Jacobsson_Studer2016_1} the authors presented a convex formulation of the problem using  the minimum mean square error (MMSE) and applied it to higher modulation scheme in \cite{Jacobsson_Studer2016_2}. Recently, \cite{Swindlehurst_2017} proposed a method to significantly improve linear precoding solutions 
in conjunction with 1-bit quantization
by properly perturbing the linearly precoded signal to favorably impact
the probability of correct detection. In this work, we provide a novel computationally efficient technique to transmit PSK symbols through a massive MIMO downlink  channel with 1-bit DACs  based on linear programming. This method is based on a distance metric for minimizing the probability of detection errors, rather than the MMSE criterion which is quite restrictive due to the finite data alphabet. The linear programming type of formulation is very advantageous in terms of complexity as it is one of the most widely applied and studied optimization technique.

This paper is organized as follows: in Section \ref{sec:sysmodel} we present the system model. In Sections \ref{sec:precoding_task} and \ref{sec:opt_problem} we introduce the 1-bit precoding problem, and formulate the design criterion and the optimization problem as a linear programming problem, respectively. In Sections \ref{sec:simresults} and \ref{sec:conclusion} we show the simulation results and summarize this work.

\textbf{Notation}: Bold lower case and upper case letters indicate vectors and matrices, non-bold letters express scalars. The operators $(.)^{*}$, $(.)^{\T}$ and $(.)^{\He}$ stand for complex conjugation, transposition and Hermitian transposition, respectively. The $n \times n$ identity (zeros) matrix is denoted by $\matt{I}_{n}$ ($\matt{0}_{n}$). The $n$ dimensional one (zero) vector is denoted by $\vect 1_n$ ($\vect 0_n$). The vector $\vect{e}_m$ represents a zero-vector with $1$ at the $m\text{th}$ position. Additionally, $\diag\left(\vect a\right)$ denotes a diagonal matrix containing the entries of the vector $\vect{a}$.
Every vector $\vect{a}$ of dimension $L$ is defined as $\vect a = \sum_{\ell=1}^L a_{\ell} \vect{e}_l$.
\section{System Model}
\label{sec:sysmodel}
\begin{figure}[h]
\centering
\scalebox{.75}{\tikzstyle{int}=[draw, minimum width=1cm, minimum height=1cm,  very thick]
\tikzstyle{init} = [pin edge={<-,thick,black}]
\tikzstyle{sum} = [draw, circle,inner sep=1pt, minimum size=2mm, very thick] 

\begin{tikzpicture}[node distance=1cm,auto,>=latex']
    \node [int] (a) {$\mathcal{P}\left(\bullet\right)$};
    \node (b) [left of=a,node distance=1.5cm, coordinate] {a};
    \node [int] (c) [right=1cm of a] {$\Q\left(\bullet\right)$};
    \node[int] (f) [right=1cm of c]  {$\sqrt{\frac{P_\text{tx}}{N}} \matt I$};
\node[int] (e) [right=0.5cm of f]  {$\matt H$};
    \node [sum,  pin={[init]below:$\boldsymbol{\eta}$}] (g) [right=1cm of e] {$\matt{+}$};
   
     \node [int] (d) [right=0.5cm of g] {$\mathcal{D}$};
    \node [coordinate] (end) [right=1cm of d, node distance=1cm]{};

    \path[->,thick] (b) edge node[above] {$\vect{s}$} 
                       node[below] {$\mathcal{O}_{D}^{M}$} (a);
    \path[->,thick] (a) edge node[above] {$\vect{x}$}
                       node[below] {$\mathbb{C}^{N}$} (c);
    \path[->,thick] (c) edge node[above] {$\vect{x}_Q$} 
                       node[below] {$\mathcal{O}_{4}^{N}$} (f);
    \path[->,thick] (f) edge  (e);                       
    \path[->,thick] (e) edge node[above] {$\vect{y}$} 
                       node[below] {$\mathbb{C}^{M}$} (g);
    \path[->,thick] (g) edge node[above] {$\vect{r}$} 
                       node[below] {$\mathbb{C}^{M}$} (d);                      
    \path[->,thick] (d) edge node[above] {$\hat{\vect{s}}$} 
                      node[below] {$\mathcal{O}_{D}^{M}$} (end) ;
\end{tikzpicture}}
\caption{Downlink quantized MU-MISO system model}
\label{fig:model}
\end{figure}
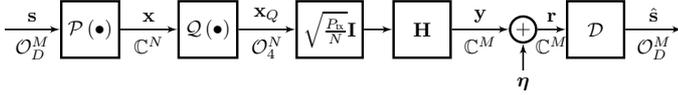
The system model shown in Fig.\ref{fig:model} consists of a massive MU-MISO downlink scenario with 1-bit quantization at the transmitter. The base station (BS) is equipped with $N$ antennas and serves $M$ single-antenna users simultaneously, where $N\gg M$.\\
The input signal $\mathbf{s}\in\mathcal{O}_D^M$ contains the symbols to be transmitted to each of the $M$ users, where
$\mathcal{O}_D$ represents the set of the $D$-PSK constellation,  We assume that $\mathrm{E}[\vect{s}]=\vect{0}_{M}$ and $\mathrm{E}[\vect{s}\vect{s}^{\He}]=\sigma^2_\text{s}\matt{I}_{M}$. The signal vector $\vect{s}$ is mapped into the vector $\vect{x}$ prior to the DAC. The precoder $\mathcal{P}$ is a symbol-wise nonlinear precoder to reduce the distortions caused by the coarse quantization and the channel distortions. The precoder output reads as
\begin{align}
\vect x = \mathcal{P}\left(\vect s, \matt H\right), 
\end{align} 
where $\matt{H}$ is the channel matrix with the $\left(m,n\right)\text{th}$ element $h_{mn}$ being the zero-mean unit-variance channel tap between the $n$th transmit antenna and the $m$th user.
The precoder output is a function of the channel matrix $\matt H$ and the input signal vector $\vect s$. The precoding task will be explained in detail in Sections \ref{sec:precoding_task}  and \ref{sec:opt_problem}.\\
The  1-bit quantization $\mathcal{Q}$ delivers then the signal vector $\vect{x}_{Q}\in\mathcal{O}_{4}^N$, where $ \mathcal{O}_{4}= \lbrace \pm \frac{1}{\sqrt{2}} \pm \jim \frac{1}{\sqrt{2}} \rbrace$. The total transmit power $P_\text{tx}$ is allocated equally among the transmit antennas, which means that the signals at each transmit antenna get scaled with $\sqrt{\frac{P_\text{tx}}{N}}$. The noiseless received signal is given by 
\begin{align}
\vect y = \sqrt{\frac{P_\text{tx}}{N}} \matt H \vect x_Q.
\end{align}
The received signal vector $\hat{\vect{s}}\in \mathcal{O}_D
^M$ after the decision block $\mathcal{D}$ can be written as follows 
\begin{align}
\hat{\vect{s}}=\mathcal{D}\left\lbrace
\sqrt{\frac{P_\text{tx}}{N}}\matt{H}\vect{x}_{Q}+\boldsymbol{\eta}\right\rbrace,
\end{align} 
where $\boldsymbol{\eta} \sim \mathcal{C} \mathcal{N}\left( \vect{0}_{M}, \mathbf{C}_{\boldsymbol{\eta}}=\matt{I}_{M}\right)$ denotes the vector of the additive white Gaussian noise (AWGN) vector at the $M$ receive antennas.

\section{Precoding Task}
\label{sec:precoding_task}
The symbol-wise precoder aims to mitigate the multi-user interference and the 1-bit quantization distortions. The task consists in designing the transmit vector $\vect x$ such that $ \hat {\vect s} = \vect s$ holds true with high probability to reduce the detection error probability.

To mitigate the quantization distortions, we design the input to the quantizer to belong to $\mathcal{O}_4^N$. Consequently, we would get in the ideal case an undistorted signal 
\begin{align}
\vect x_Q = \vect x \text{, if } \vect x \in \mathcal{O}_4^N.
\label{eq:linear_quantizer}
\end{align}
We denote (\ref{eq:linear_quantizer}) by the 1-bit constraint, that ensures the non-distorting behavior of the 1-bit quantizer $\mathcal{Q}$. This constraint, however, leads to a discrete optimization problem that cannot be solved efficiently. Therefore, the 1-bit constraint will be relaxed to a convex constraint as shown in Section \ref{sec:relaxed_constraint}. The constraint relaxation does not satisfy the equality in (\ref{eq:linear_quantizer}) and thus the quantization distortions are not fully omitted. However, they are reduced significantly as shown later.

For the next derivations, we introduce the following signal vector 
\begin{align}
\vect y' =  \vect y \mid_{P_\text{tx} =N, (\ref{eq:linear_quantizer})}= \matt H \vect x.
\label{eq:y'}
\end{align} 
This signal vector $\vect y'$ is equal to the noiseless received signal $\vect y$ for a transmit power $P_\text{tx} =N$ and when (\ref{eq:linear_quantizer}) is fulfilled. The optimization is based on this special case, since the transmit power just scales the noiseless received signal and the constraint in (\ref{eq:linear_quantizer}) is approximated with a convex constraint, that will be introduced in Section \ref{sec:relaxed_constraint}.
\section{Problem Formulation for PSK Signaling}
\label{sec:opt_problem}
\subsection{Constructive Interference Optimization}
\begin{figure}
\centering  
\psfrag{Re}[][]{$\Re$}
\psfrag{Im}[][]{$\jim \Im$}
\psfrag{theta}[][]{$\theta$}
\psfrag{s}[][]{$\:\:\:\:\:\:\:\:\:\:\:\:\:\:\:\:\:\:\:\:\:\:\:\:\:s_m$}
\psfrag{y}[][]{$\:\:\:\:\:\:\:\:\:\:\:\:\:\:\:\:\:\:\:\:\:\:\:\:\:\:\:\:\:\:\:\:\:y_m'$}
\psfrag{d1}[][]{$\!\!\!\!\!\!\!\!\!\!\!\!\!\!\!\!\!\!\!\!\!\!\!\!\!\!\!\!\!\!\!\!\!\!\!\!\!\!\!\delta$}
\psfrag{d2}[][]{$\:\:\:\:\:\:\:\:\:\:\:\:\:\:\:\:\:\:\:\:\:\:\delta$}
\psfrag{A}[][]{$\:\:\:\:\:\:\:\:\:\:\:\:\:\:\:\:\:\:\:\:\:\:\:\:\:\:\:\:\:\:\:\:\:\:\:\:\:\:\:\:\:\textcolor{green}{z_{m_R}}$}
\psfrag{B}[][]{$\:\:\:\:\:\:\:\:\:\:\:\:\:\:\:\:\:\:\:\:\:\:\:\:\:\:\:\:\:\:\:\:\:\:\:\:\:\:\:\:\:\:\:\:\:\:\:\:\:\:\:\:\:\textcolor{red}{z_{m_I}}$}
\includegraphics[width=0.3\textwidth]{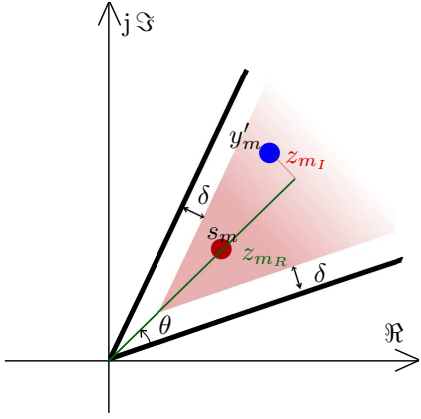}
\caption{Illustration of the symbol region.}
\label{fig:opt_region_psk}
\end{figure}

When the downlink channel and all user's data are known at the transmitter, the instantaneous constructive multi-user interference can be exploited to move the received signals further far from the decision thresholds \cite{Masouros2015}. In contrast to this, conventional precoding methods (MMSE, Zero-forcing) aim at minimizing the total multi-user interference such that the received signals lie as closed as possible to the nominal constellation points. In fact, the constructive interference optimization exploits the larger symbol decision regions and thus leads to a more relaxed optimization. Each symbol region (SR), as shown in Fig. \ref{fig:opt_region_psk}, is a circle sector of infinite radius and angle of $2\theta$, where
\begin{align}
\theta = \frac{\pi}{D}.
\end{align}
The symbol region is shifted from the decision thresholds by a safety margin denoted by $\delta$. This safety margin has to be maximized to make sure that the received symbols when perturbed by the additive noise do not jump to the neighboring unintended symbol regions.

The optimization problem can be written in general as follows
\begin{align}
&\max_{\vect x}  \delta \label{eq:max_delta} \\
\text{ s.t.  }&  y'_m \in \text{SR}, \forall m \label{eq:OR_constraint}\\
\text{and  }& \vect x \in \mathcal{O}_4^N \label{eq:1bit_constraint}.
\end{align}
In the next sections a mathematical expression for the SR is derived. In addition, the 1-bit constraint in (\ref{eq:1bit_constraint}) is relaxed to get a convex solution set.

\subsection{Symbol Regions (SR)}
\begin{figure}
\centering  
\psfrag{teta}[][]{$\:\:\:\:\:\:\:\:\:\:\:\:\:\:\:\:\:\:\:\:\:\:\:\:\:\:\:\:\:\:\:\:\:\:\theta$}
\psfrag{tau}[][]{$\:\:\:\:\:\:\tau$}
\psfrag{A}[][]{$\!\!\!\!\!\!\!\!\!\!\!\!\!\!\!\!\!\!\!\!\!\!\!\!\!\!\!\!\!\!\!\!\!\!\!\!\!z_{m_R}$}
\psfrag{B}[][]{$\:\:\:\:\:z_{m_I}$}
\includegraphics[width=0.3\textwidth]{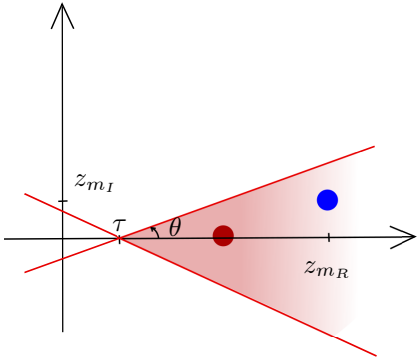}
\caption{Illustration of the symbol region in a modified coordinates system.}
\label{fig:opt_region_psk_mod_coordinates}
\end{figure}

To determine the SR, a modified coordinates system is considered as illustrated in Fig. \ref{fig:opt_region_psk_mod_coordinates}.
The coordinates system is rotated by the phase of the symbol of interest $s_m$. The coordinates of the noiseless received signal $y'_m$ in the modified coordinates system are given by
\begin{align}
z_{m_R} &= \Re\lbrace y_m' s_m^*\rbrace \frac{1}{\vert s_m\vert }, \\
z_{m_I} &= \Im\lbrace y_m' s_m^*\rbrace \frac{1}{\vert s_m\vert } .
\end{align}
Since PSK signals have unit magnitude, plugging in (\ref{eq:y'}) into the above equations gives
\begin{align}
z_{m_R} &= \Re\lbrace \vect e_m^{\T} \matt H \vect x s_m^*\rbrace = \Re\lbrace \vect e_m^{\T} \matt {\tilde H} \vect x \rbrace, \label{eq:z_R} \\
z_{m_I} &= \Im\lbrace \vect e_m^{\T} \matt H \vect x s_m^*\rbrace = \Im\lbrace \vect e_m^{\T} \matt {\tilde H} \vect x \rbrace,\label{eq:z_I}
\end{align}
where $\matt {\tilde H} = \diag(\vect s^*) \matt H$. We define $\matt {\tilde H}$ as the modified channel.
The symbol region can be hence described by
\begin{align}
&z_{m_R} \geq \tau \label{eq:z_R_constraint} \\
&\vert z_{m_I} \vert \leq \left( z_{m_R} - \tau \right) \tan{\theta}, \forall m, \label{eq:z_I_constraint}
\end{align}
where $\tau = \frac{\delta}{\sin{\theta}}$. Note that the inequality in (\ref{eq:z_R_constraint}) is already fulfilled if the inequality in (\ref{eq:z_I_constraint}) is satisfied.
Plugging in (\ref{eq:z_R}) and (\ref{eq:z_I}) into (\ref{eq:z_I_constraint}), the symbol regions for all $M$ users can be defined by
\begin{align}
\vert \Im\lbrace \matt {\tilde H} \vect x \rbrace\vert \leq \left( \Re\lbrace \matt {\tilde H}\vect x \rbrace - \tau \vect 1_M\right) \tan{\theta}.
\end{align}
When using the following real-valued representation
\begin{align}
\Re\lbrace \matt {\tilde H}\vect x \rbrace &= \underbrace{\begin{bmatrix}
\Re\lbrace \matt {\tilde H} \rbrace & -\Im\lbrace \matt {\tilde H} \rbrace \end{bmatrix}}_{=\matt A} \underbrace{\begin{bmatrix} \Re\lbrace \vect x \rbrace\\ \Im\lbrace \vect x \rbrace
\end{bmatrix}}_{= \vect x'}= \matt A \vect x' \\
\Im\lbrace \matt {\tilde H}\vect x \rbrace &= \underbrace{\begin{bmatrix}
\Im\lbrace \matt {\tilde H} \rbrace & \Re\lbrace \matt {\tilde H} \rbrace \end{bmatrix}}_{=\matt B} \begin{bmatrix} \Re\lbrace \vect x \rbrace\\ \Im\lbrace \vect x \rbrace
\end{bmatrix} = \matt B \vect x',
\end{align}
the constraint in (\ref{eq:OR_constraint}) can be rewritten as
\begin{align}
\begin{bmatrix} \matt B - \tan{\theta}\matt A &
    \frac{1}{\cos{\theta}} \vect 1_M \\ -\matt B - \tan{\theta}\matt A &
    \frac{1}{\cos{\theta}} \vect 1_M\end{bmatrix} \begin{bmatrix}\vect x' \\ \delta \end{bmatrix}\leq \vect 0_{2M}.
    \label{eq:OR_linear_constraint}
\end{align}

\subsection{Relaxed 1-Bit Constraint}
\label{sec:relaxed_constraint}
The 1-bit constraint, $\vect x \in \mathcal{O}_4^N$ makes sure that the quantization distortions are avoided since it leads to $\vect x_Q = \vect x$. However, this constraint is non-convex. Thus, the constraint is relaxed such that the entries of the vector $\vect x$ belong to the filled box built by the QPSK symbols.
We can describe the relaxed convex constraint as follows
\begin{align}
\vect x' \leq \frac{1}{\sqrt{2}}\vect 1_{2N} \text{ and }
-\vect x' \leq \frac{1}{\sqrt{2}} \vect 1_{2N}.
\end{align} 
Hence, the constraint in (\ref{eq:1bit_constraint}) is replaced by the following relaxed convex constraint
\begin{align}
\begin{bmatrix}
\matt I_{2N} & \vect 0_{2N} \\
-\matt I_{2N} & \vect 0_{2N}
\end{bmatrix} \begin{bmatrix}\vect x' \\ \delta \end{bmatrix}\leq \frac{1}{\sqrt{2}}\vect 1_{4N}.
\label{eq:relaxed_1bit_constraint}
\end{align}
\subsection{Optimization Problem with the Relaxed Constraint}
Combining (\ref{eq:max_delta}), (\ref{eq:OR_linear_constraint}) and (\ref{eq:relaxed_1bit_constraint}), the optimization problem can be finally expressed by the following real-valued linear programming problem
\begin{align}
&\max_{\vect v} \begin{bmatrix}
\vect 0_{2N}^{\T} & 1
\end{bmatrix} \vect v \nonumber \\
&\text{ s.t. }  
\begin{bmatrix} \matt B - \tan{\theta}\matt A &
    \frac{1}{\cos{\theta}} \vect 1_M \\ -\matt B - \tan{\theta}\matt A &
    \frac{1}{\cos{\theta}} \vect 1_M \\ 
    \matt I_{2N} & \vect 0_{2N} \\
    -\matt I_{2N} & \vect 0_{2N}  \end{bmatrix} \vect v \leq \begin{bmatrix} \vect 0_{2M} \\ \frac{1}{\sqrt{2}} \vect 1_{4N} \end{bmatrix},
    \label{eq:final_optimization_problem}
\end{align}
where $\vect v^{\T} = \begin{bmatrix}
\vect x'^{\T} & \delta
\end{bmatrix}$. This linear programming problem has $\left(2N+1\right)$ unknowns and $\left(2M+4N\right)$ inequalities to satisfy.
\section{Complexity}
The linear programming is a very popular convex optimization technique, that is efficiently  solvable using a wide variety of methods \cite{Boyd2004}.
With the use of interior-point methods the number of iterations almost always lies between 10 and 100 \cite{Boyd2004}. Each iteration requires a number of arithmetic operations on the order of 
\begin{align}
c &=\max \lbrace (2N+1)^3,(2N+1)^2(2M+4N),4NM\rbrace\nonumber \\
 &= (2N+1)^2(2M+4N).
\end{align}
The complexity of linear programming is therefore bounded, which makes it attractive for hardware implementation.
\section{Simulation Results}
\label{sec:simresults}
For the simulations, we assume a
BS with $N = 128$ antennas serving $M = 16$ single-antenna
users. The channel $\matt H$ is composed of i.i.d. Gaussian
random variables with zero-mean and unit variance. 
All the simulation results are obtained with $N_b=10^3$ transmit symbols per channel use with $\sigma_s^2=1$ and averaged over 100 channel realizations. 
The additive
noise is also i.i.d with variance one at each antenna. 
The performance metric is the uncoded BER. The considered modulation schemes are QPSK, 8PSK and 16 PSK. We compare our proposed design maximum safety margin (MSM) with the linear precoder quantized Wiener Filter (QWF) from \cite{Mezghani2009}, the symbol-wise precoder $\text{SDR}_{\ell^2_{\infty}}$ in (47) from\cite{Jacobsson_Studer2016_2}, the symbol-wise precoder MBER from \cite{JeddaSAM2016} and the ideal case denoted by "WF, unq.", where the WF precoder is used and no quantization is performed. The MBER precoder is restricted to QPSK symbols. The comparison is conducted for two scenarios
\begin{itemize}
\item perfect CSI and
\item imperfect CSI.
\end{itemize}

Assuming full CSI, the uncoded BER is plotted as function of the available transmit power $P_\text{tx}$ for three modulation schemes: QPSK (Fig. \ref{fig:ber_128_16_4PSK}), 8 PSK (Fig. \ref{fig:ber_128_16_8PSK}) and 16 PSK (Fig. \ref{fig:ber_128_16_16PSK}). It can be seen from the results that the proposed precoder MSM outperforms the linear precoder QWF and the MBER precoder (for QPSK). The gain in dB compared to QWF increases when the order modulation increases. However, for higher order PSK modulation, the proposed symbol-wise precoder still presents an error floor, meaning that potentially higher number of antennas are needed in this case.\\
The loss due to the 1-bit quantization compared to the ideal case "WF, unq." increases with higher order modulation, 2dB, 3dB and 6dB at BER of $10^{-2}$ for QPSK, 8 PSK and 16 PSK, repectively.\\
The MSM precoder performs almost the same as the $\text{SDR}_{\ell^2_{\infty}}$ precoder. However, the complexity of the proposed method is very low as the simulations show that only 14 iterations in average are needed to solve (\ref{eq:final_optimization_problem}).

\begin{figure}
\centering
\resizebox{9cm}{!} {
%
%

\definecolor{mycolor1}{rgb}{0.00000,0.44700,0.74100}%
\begin{tikzpicture}

\begin{axis}[%
width=4.521in,
height=2.5in,
at={(0.758in,0.481in)},
scale only axis,
xmin=0,
xmax=24,
xlabel={$P_{\text{tx}}$ (dB)},
xmajorgrids,
ymode=log,
ymin=1e-04,
ymax=0,
yminorticks=true,
ylabel={Uncoded BER},
ymajorgrids,
yminorgrids,
axis background/.style={fill=white},
title style={font=\bfseries},
legend style={legend cell align=left,align=left,draw=white!15!black}
]
\addplot [color=blue,solid,line width=2.0pt]
  table[row sep=crcr]{%
0	0.02228\\
2	0.0063753125\\
4	0.001058125\\
6	9.31250000000001e-05\\
8	4.375e-06\\
10	0\\
12	0\\
14	0\\
16	0\\
18	0\\
20	0\\
22	0\\
24	0\\
};
\addlegendentry{MSM, Full CSI};

\addplot [color=blue,solid,line width=2.0pt,mark=square,mark options={solid}]
  table[row sep=crcr]{%
0	0.0233840625\\
2	0.006979375\\
4	0.0012509375\\
6	0.000125\\
8	5.9375e-06\\
10	0\\
12	0\\
14	0\\
16	0\\
18	0\\
20	0\\
22	0\\
24	0\\
};
\addlegendentry{MSM, $\upsilon^2=0.1$};

\addplot [color=blue,solid,line width=2.0pt,mark=x,mark options={solid}]
  table[row sep=crcr]{%
0	0.026466875\\
2	0.0086309375\\
4	0.0019215625\\
6	0.0002771875\\
8	2.5625e-05\\
10	1.25e-06\\
12	0\\
14	0\\
16	3.125e-07\\
18	0\\
20	0\\
22	0\\
24	0\\
};
\addlegendentry{MSM, $\upsilon^2=0.2$};

\addplot [color=red,solid,line width=2.0pt]
  table[row sep=crcr]{%
0	0.0942727083333333\\
2	0.0590647916666667\\
4	0.0315522916666667\\
6	0.0136414583333333\\
8	0.00461854166666667\\
10	0.001200625\\
12	0.00028375\\
14	7.04166666666667e-05\\
16	1.95833333333334e-05\\
18	7.91666666666666e-06\\
20	6.875e-06\\
22	4.79166666666667e-06\\
24	4.58333333333333e-06\\
};

\addplot [color=red,solid,line width=2.0pt,mark=square,mark options={solid}]
  table[row sep=crcr]{%
0	0.0960879166666666\\
2	0.0607497916666667\\
4	0.0333414583333333\\
6	0.0150308333333333\\
8	0.005528125\\
10	0.001666875\\
12	0.000457083333333333\\
14	0.000125833333333333\\
16	4.66666666666667e-05\\
18	2.27083333333334e-05\\
20	1.45833333333333e-05\\
22	1.04166666666667e-05\\
24	8.95833333333333e-06\\
};

\addplot [color=red,solid,line width=2.0pt,mark=x,mark options={solid}]
  table[row sep=crcr]{%
0	0.101057083333333\\
2	0.0659539583333333\\
4	0.0383847916666667\\
6	0.0195016666666667\\
8	0.0087425\\
10	0.00354125\\
12	0.00142708333333333\\
14	0.00061125\\
16	0.000320625\\
18	0.000188125\\
20	0.000135\\
22	0.000106875\\
24	7.97916666666667e-05\\
};

\addplot [color=green,solid,line width=2.0pt,forget plot]
  table[row sep=crcr]{%
0	0.1731196875\\
2	0.1349228125\\
4	0.10232\\
6	0.07490015625\\
8	0.05201984375\\
10	0.03395265625\\
12	0.0209271875\\
14	0.012900625\\
16	0.0081809375\\
18	0.00560125\\
20	0.00421109375\\
22	0.0034225\\
24	0.00299359375\\
};

\addplot  [color=green,solid,line width=2.0pt,mark=square,mark options={solid}]
  table[row sep=crcr]{%
0	0.1748990625\\
2	0.136969375\\
4	0.10443203125\\
6	0.0775309375\\
8	0.05516765625\\
10	0.0375071875\\
12	0.02465671875\\
14	0.01642390625\\
16	0.0113153125\\
18	0.0083496875\\
20	0.00670109375\\
22	0.00570203125\\
24	0.0051096875\\
};

\addplot [color=green,solid,line width=2.0pt,mark=x,mark options={solid}]
  table[row sep=crcr]{%
0	0.18007171875\\
2	0.14268859375\\
4	0.11091390625\\
6	0.0849696874999999\\
8	0.06422453125\\
10	0.04776078125\\
12	0.0356565625\\
14	0.027461875\\
16	0.02201390625\\
18	0.0185421875\\
20	0.0164015625\\
22	0.01507984375\\
24	0.01420953125\\
};
\end{axis}
\end{tikzpicture}%
}
\caption{BER performance for a MU-MISO system with $N=128$ and $M=16$ with imperfect CSI: QPSK (blue), 8 PSK (red) and 16 PSK (green).}
\label{fig:ber_128_16_4PSK_ch_est}
\end{figure}
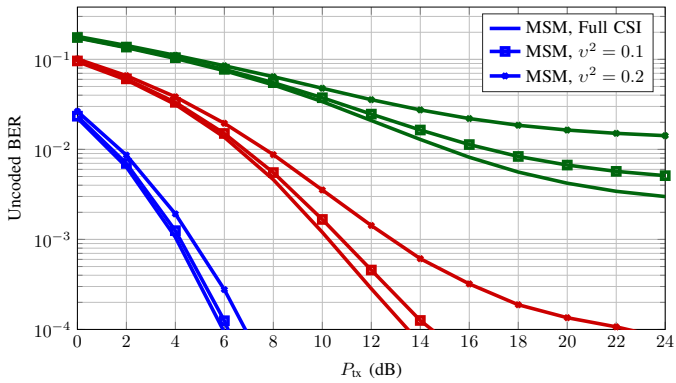

Next, the effect of imperfect CSI on the performance of the proposed precoder is considered. To this end, we assume that the channel matrix $\matt H$ is perturbed with an error matrix $\matt \Gamma$, that has i.i.d. entries with zero mean and variance $\upsilon^2$. The channel estimate can be then written as
\begin{align}
\matt H_e = \matt H +\matt {\Gamma}.
\end{align}
The optimization problem in (\ref{eq:final_optimization_problem}) is then run with $\matt H_e$ for $\upsilon^2=0$ (full CSI), $\upsilon^2=0.1$ and $\upsilon^2=0.2$ for the three modulation schemes. The simulation results are shown in Fig. \ref{fig:ber_128_16_4PSK_ch_est}. As can be concluded, the MSM precoder is more robust against imperfect CSI for QPSK. For higher order modulation, the loss due to channel estimation errors increases.

\begin{figure}
\centering
\resizebox{9cm}{!} {
%
%
\definecolor{mycolor1}{rgb}{0.00000,0.44700,0.74100}%
\begin{tikzpicture}

\begin{axis}[%
width=4.521in,
height=2.5in,
at={(0.758in,0.481in)},
scale only axis,
xmin=0,
xmax=24,
xlabel={$P_{\text{tx}}$ (dB)},
xmajorgrids,
ymode=log,
ymin=1e-04,
ymax=0.1,
yminorticks=true,
ylabel={Uncoded BER},
ymajorgrids,
yminorgrids,
axis background/.style={fill=white},
title style={font=\bfseries},
legend style={legend cell align=left,align=left,draw=white!15!black}
]

\addplot [color=blue,solid,line width=2.0pt]
  table[row sep=crcr]{%
0	0.02228\\
2	0.0063753125\\
4	0.001058125\\
6	9.31250000000001e-05\\
8	4.375e-06\\
10	0\\
12	0\\
14	0\\
16	0\\
18	0\\
20	0\\
22	0\\
24	0\\
};
\addlegendentry{MSM};

\addplot [color=blue,solid,line width=2.0pt,mark=+,mark options={mark size=3,solid}]
  table[row sep=crcr]{%
0	0.0224796875\\
2	0.0064271875\\
4	0.001053125\\
6	9.62500000000001e-05\\
8	4.6875e-06\\
10	0\\
12	0\\
14	0\\
16	0\\
18	0\\
20	0\\
22	0\\
24	0\\
};
\addlegendentry{$\text{SDR}_{\ell^2_{\infty}}$};

\addplot [color=blue,solid,line width=2.0pt,mark=triangle,mark options={mark size=3,solid,rotate=180}]
  table[row sep=crcr]{%
0	0.02395426875\\
2	0.00867668124999999\\
4	0.002365525\\
6	0.000466825\\
8	6.74812499999999e-05\\
10	7.13125000000002e-06\\
12	5.875e-07\\
14	9.375e-08\\
16	3.75e-08\\
18	1.25e-08\\
20	1.25e-08\\
22	1.25e-08\\
24	1.875e-08\\
};
\addlegendentry{MBER};

\addplot [color=red,solid,line width=2.0pt,mark=o,mark options={mark size=3,solid}]
  table[row sep=crcr]{%
0	0.03205388125\\
2	0.01527884375\\
4	0.00679144375000001\\
6	0.00299815\\
8	0.00142105\\
10	0.00076043125\\
12	0.00046765625\\
14	0.00032814375\\
16	0.00025810625\\
18	0.00021836875\\
20	0.00019385625\\
22	0.00018025625\\
24	0.0001723625\\
};
\addlegendentry{QWF};

\addplot [color=red,dashed,line width=2.0pt,mark=+,mark options={mark size=3,solid}]
  table[row sep=crcr]{%
0	0.00375625\\
2	0.00041375\\
4	1.125e-05\\
6	0\\
8	0\\
10	0\\
12	0\\
14	0\\
16	0\\
18	0\\
20	0\\
22	0\\
24	0\\
};
\addlegendentry{WF, unq.};

\end{axis}
\end{tikzpicture}%
}
\caption{BER performance for a MU-MISO system with $N=128$ and $M=16$ with QPSK signaling.}
\label{fig:ber_128_16_4PSK}
\end{figure}
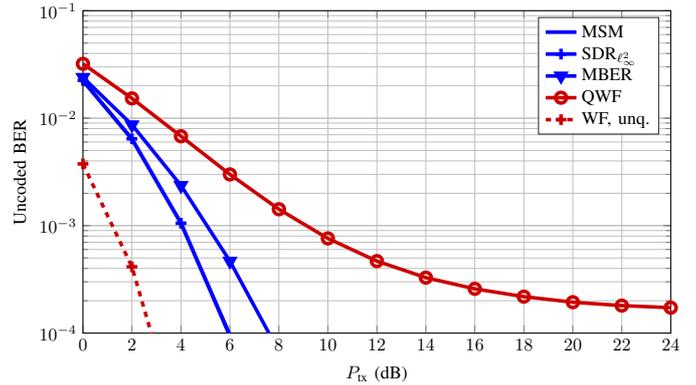

\begin{figure}
\centering
\resizebox{9cm}{!} {
%
%
\begin{tikzpicture}

\begin{axis}[%
width=4.521in,
height=2.5in,
at={(0.758in,0.481in)},
scale only axis,
xmin=0,
xmax=24,
xlabel={$P_{\text{tx}}$ (dB)},
xmajorgrids,
ymode=log,
ymin=1e-04,
ymax=1,
yminorticks=true,
ylabel={Uncoded BER},
ymajorgrids,
yminorgrids,
axis background/.style={fill=white},
title style={font=\bfseries},
legend style={legend cell align=left,align=left,draw=white!15!black}
]

\addplot [color=blue,solid,line width=2.0pt]
  table[row sep=crcr]{%
0	0.0942727083333333\\
2	0.0590647916666667\\
4	0.0315522916666667\\
6	0.0136414583333333\\
8	0.00461854166666667\\
10	0.001200625\\
12	0.00028375\\
14	7.04166666666667e-05\\
16	1.95833333333334e-05\\
18	7.91666666666666e-06\\
20	6.875e-06\\
22	4.79166666666667e-06\\
24	4.58333333333333e-06\\
};
\addlegendentry{MSM};

\addplot [color=blue,solid,line width=2.0pt,mark=+,mark options={mark size=3,solid}]
  table[row sep=crcr]{%
0	0.094515625\\
2	0.05909875\\
4	0.0314977083333333\\
6	0.0136602083333333\\
8	0.004545\\
10	0.00114708333333333\\
12	0.000255833333333333\\
14	5.27083333333333e-05\\
16	1.08333333333333e-05\\
18	3.75e-06\\
20	1.45833333333333e-06\\
22	1.04166666666667e-06\\
24	1.04166666666667e-06\\
};
\addlegendentry{$\text{SDR}_{\ell^2_{\infty}}$};

\addplot [color=red,solid,line width=2.0pt,mark=o,mark options={mark size=3,solid}]
  table[row sep=crcr]{%
0	0.109814333333333\\
2	0.0820126124999999\\
4	0.060898725\\
6	0.0457129958333334\\
8	0.0353497166666666\\
10	0.0285720583333333\\
12	0.0242283791666667\\
14	0.0214985708333333\\
16	0.0197968583333333\\
18	0.01872995\\
20	0.018056325\\
22	0.0176202833333333\\
24	0.0173571916666667\\
};
\addlegendentry{QWF};

\addplot [color=red,dashed,line width=2.0pt,mark=+,mark options={mark size=3,solid}]
  table[row sep=crcr]{%
0	0.0493764583333333\\
2	0.023125625\\
4	0.00753895833333333\\
6	0.00138979166666667\\
8	0.000104375\\
10	1.66666666666667e-06\\
12	0\\
14	0\\
16	0\\
18	0\\
20	0\\
22	0\\
24	0\\
};
\addlegendentry{WF, unq.};

\end{axis}
\end{tikzpicture}%
}
\caption{BER performance for a MU-MISO system with $N=128$ and $M=16$ with 8 PSK signaling.}
\label{fig:ber_128_16_8PSK}
\end{figure}
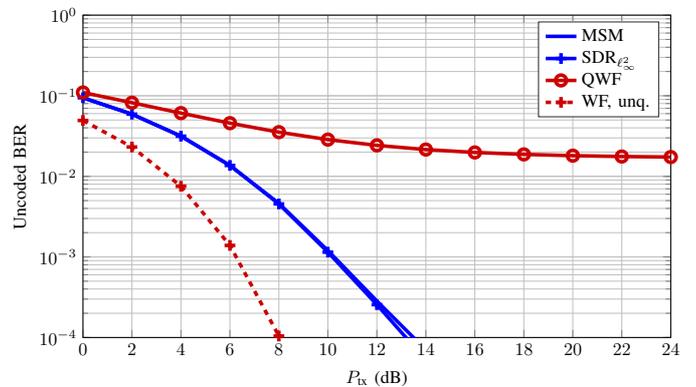

\begin{figure}
\centering
\resizebox{9cm}{!} {
%
%
\begin{tikzpicture}

\begin{axis}[%
width=4.521in,
height=2.5in,
at={(0.758in,0.481in)},
scale only axis,
xmin=0,
xmax=24,
xlabel={$P_{\text{tx}}$ (dB)},
xmajorgrids,
ymode=log,
ymin=1e-04,
ymax=1,
yminorticks=true,
ylabel={Uncoded BER},
ymajorgrids,
yminorgrids,
axis background/.style={fill=white},
title style={font=\bfseries},
legend style={at={(0.97,0.03)},anchor=south east,legend cell align=left,align=left,draw=white!15!black}
]

\addplot [color=blue,solid,line width=2.0pt]
  table[row sep=crcr]{%
0	0.1731196875\\
2	0.1349228125\\
4	0.10232\\
6	0.07490015625\\
8	0.05201984375\\
10	0.03395265625\\
12	0.0209271875\\
14	0.012900625\\
16	0.0081809375\\
18	0.00560125\\
20	0.00421109375\\
22	0.0034225\\
24	0.00299359375\\
};
\addlegendentry{MSM};

\addplot [color=blue,solid,line width=2.0pt,mark=+,mark options={mark size=3,solid}]
  table[row sep=crcr]{%
0	0.17332828125\\
2	0.13496375\\
4	0.102245\\
6	0.07462984375\\
8	0.051625\\
10	0.033495625\\
12	0.02038703125\\
14	0.0123575\\
16	0.00764140625\\
18	0.0050346875\\
20	0.00364484375\\
22	0.00286421875\\
24	0.002410625\\
};
\addlegendentry{$\text{SDR}_{\ell^2_{\infty}}$};

\addplot [color=red,solid,line width=2.0pt,mark=o,mark options={mark size=3,solid}]
  table[row sep=crcr]{%
0	0.188989040625\\
2	0.16006955625\\
4	0.1370530375\\
6	0.11972355625\\
8	0.1071840875\\
10	0.09836728125\\
12	0.0923207125\\
14	0.08826693125\\
16	0.0856128125\\
18	0.08388944375\\
20	0.0827837\\
22	0.0820665187500001\\
24	0.081624028125\\
};
\addlegendentry{QWF};

\addplot [color=red,dashed,line width=2.0pt,mark=+,mark options={mark size=3,solid}]
  table[row sep=crcr]{%
0	0.124031875\\
2	0.09077921875\\
4	0.06147015625\\
6	0.03590828125\\
8	0.01646453125\\
10	0.0051609375\\
12	0.00091640625\\
14	5.90625e-05\\
16	4.6875e-07\\
18	0\\
20	0\\
22	0\\
24	0\\
};
\addlegendentry{WF, unq.};

\end{axis}
\end{tikzpicture}%
}
\caption{BER performance for a MU-MISO system with $N=128$ and $M=16$ with 16 PSK signaling.}
\label{fig:ber_128_16_16PSK}
\end{figure}
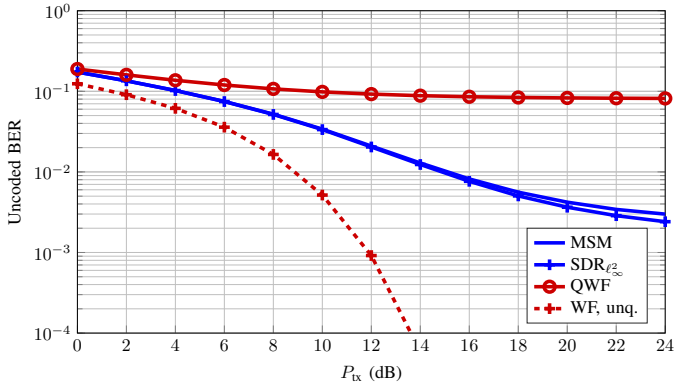

\section{Conclusion}
\label{sec:conclusion}
We proposed a symbol-wise precoder to transmit PSK signals in MU-MISO systems when 1-bit quantization is applied at the transmitter. The design of the transmit vector is based on maximizing the safety margin to the decision thresholds of the PSK modulation. The 1-bit constraint is relaxed to the box constraint and we end up with a linear programming problem that can be efficiently solved. The simulation results show a significant improvement compared to the linear precoder QWF \cite{Mezghani2009}. The proposed method performs almost the same as the method presented in \cite{Jacobsson_Studer2016_2} but with very low complexity.

\bibliographystyle{IEEEtran}
\bibliography{refs}

\end{document}